\documentclass[prl,twocolumn,showpacs]{revtex4}

\usepackage{graphicx}
\usepackage{amsmath}
\usepackage{amssymb}
\usepackage{epsf,latexsym}
\usepackage{times}

\newcommand{\ket}[1]{\left| #1 \right\rangle}
\newcommand{\bra}[1]{\left\langle #1 \right|}
\newcommand{\be}{\begin{equation}}
\newcommand{\ee}{\end{equation}}
\newcommand{\ba}{\begin{eqnarray}}
\newcommand{\ea}{\end{eqnarray}}

\begin{document}


\title{Freezing distributed entanglement in spin chains}

\author{Irene D'Amico$^{1}$}
\email{ida500@york.ac.uk}
\author{Brendon W. Lovett$^{2}$}
\author{Timothy P. Spiller$^{3}$}

\affiliation{
$^1$ Department of Physics, University of York, York YO10 5DD, United Kingdom\\
$^2$ Department of Materials, University of Oxford, OX1 3PH, United Kingdom\\
$^3$ Hewlett-Packard Laboratories, Filton Road,
Stoke Gifford, Bristol BS34 8QZ, United Kingdom}

\date{\today }

\begin{abstract}
We show how to freeze distributed entanglement that has been created from the
natural dynamics of spin chain systems. The technique that we propose simply requires single-qubit operations and isolates the
entanglement in specific qubits at the ends of branches.
Such frozen entanglement provides a
useful resource, for example for teleportation or distributed
quantum processing. The scheme can be applied to a wide range of systems -- including actual spin systems and alternative qubit embodiments in strings of quantum dots,
molecules or atoms.
\end{abstract}



\maketitle

Over the past few years there has been significant interest in the propagation of quantum information through spin chains, which has been stimulated by potential realizations in systems such as large molecules, quantum dot arrays and trapped atoms.
It is fundamentally
interesting to determine just what the natural dynamics of these
systems permits -- and from a practical point of view the controlled propagation of quantum information
is likely to be an essential part of any quantum computing device. For macroscopic distances quantum states
of light are almost certain to form the best medium for quantum communication -- but
for shorter, microscopic, distances it is equally likely that other
media, such as spin chains, could make a very useful contribution.
For example, in future solid state devices there could be a need to
provide quantum communication links over microscopic distances,
between separate quantum processors or registers, or between
processors and memory, analogous to the conventional communication
that goes on within the computer chips of today.

Distributed entangled states can be used in order to improve the fidelity of a quantum state transmission, which then proceeds through teleportation. The entanglement can be prepared off line, and then purified \cite{pur96} prior to
its application. A distributed resource can also facilitate quantum repetition \cite{repeater98},
quantum gates between separated systems or the construction of
extended entangled resources such as cluster states \cite{cluster}.
Previous work has focussed on the production and distribution of entanglement
but, in order to use it as a resource, it is essential that the entanglement can be stored or frozen for future operations.
In this Letter we present a technique for doing exactly this,
using the natural dynamics of spin chain systems.
We will use `spin chain' to cover any
physical system whose dynamics can be predicted using a Hamiltonian
isomorphic to that of a coupled spin chain. This could
be strings of actual spins (produced chemically, or fabricated)
connected through interactions, or it could be a string of quantum
dots or molecules (like fullerenes), containing exciton or spin
qubits. Strings of trapped atoms form another possibility.

Production of a useful, frozen, entangled resource requires
production and good fidelity distribution of entanglement, followed
by intervention to freeze it. There has been significant study of
the propagation of quantum states through spin chains or networks.
Originally it was shown that a single spin qubit could transfer with
decent fidelity along a constant nearest-neighbour exchange-coupled
chain \cite{bose03}. The fidelity can approach unity if the qubit is
encoded into a packet of spins \cite{osb04}. Perfect state transfer
can also be achieved in more complicated systems, with different
geometry or chosen unequal couplings
\cite{chr04,chr05}. Parallel spins chains also
enable perfect transfer \cite{bur05}, as does a chain used
as a wire, with controlled couplings at the ends \cite{woj05}.
Related studies have also been made on chains of quantum dots
\cite{damico05,spi06}, chains of quantum oscillators
\cite{ple05} and spin chains connected through long range
magnetic dipole interactions \cite{avel06}. State transfer and
operations through spin chains using adiabatic dark passage has also
been proposed \cite{gre05}. We will discuss the creation of
high fidelity spatially separated entanglement---in a form suitable
for freezing---from the natural dynamics of branched spin chain
systems, and then show how it can be frozen. The consequences of
branching were first studied in the context of divided {\it bosonic}
chains, composed of coupled harmonic oscillators~\cite{plenio04,
perales05} - and have since been studied in systems of propagating
electrons~\cite{yang06}. In both of these cases, the
dynamics of Gaussian wave packet type excitations have been
investigated. Our work deals with the propagation of a single
spin-down excitation localized on a single site, and how this moves
through a network of spin-up states

To introduce our formalism, we first consider a one-dimensional
chain of $N$ spins, each coupled to their nearest neighbours. The
Hamiltonian for the system is
\be
H = -\sum_{i=1}^N \frac{E_i}{2} \sigma_{z}^{i}+
\sum_{i=1}^{N-1} \frac{J_{i,i+1}}{2}\left(\sigma_{+}^{i}\sigma_{-}^{i+1} + \sigma_{-}^{i}\sigma_{+}^{i+1}\right)
\;,
\label{Hspin}
\ee
where $\sigma_{z}^{i}$ is the $z$ Pauli spin matrix for the spin
at site $i$ and similarly $\sigma_{\pm}^i=\sigma_{x}^{i} \pm i
\sigma_{y}^{i}$. For actual spins, $E_{i}/2$ is the local magnetic
field (in the $z$-direction) at site $i$ and $J_{i,i+1}/2$ is the
local $XY$ coupling strength between neighbouring sites $i$ and
$i+1$. For a coupled chain of quantum dots where each qubit is
represented by the presence or absence of a ground state exciton,
$E_i$ is the exciton energy and $J_{i,i+1}$ is the F\"orster
coupling between dots $i$ and $i+1$ \cite{damico05}. The spin chain
formalism applies to both such physical systems, and to any others
which can be described by an isomorphic Hamiltonian. The
computational basis notation for the spin states at each site is
$|0\rangle_i \equiv |\uparrow_z\rangle_i$ and $|1\rangle_i \equiv
|\downarrow_z\rangle_i$. The total $z$-component of spin
(magnetization), or total exciton number, is a constant of motion as
it commutes with $H$. It is therefore instructive to consider a
state of the system consisting of the ground state with the addition
of a single flipped spin. This state is straightforward to prepare,
assuming local control over a spin at, say, the end of a chain, and
the flipped spin can be regarding as a (conserved) travelling qubit
as it moves around under the action of the chain dynamics
\cite{bose03}. An efficient way of representing such states in an
$N$ spin network is the site basis defined as $\ket{\mathsf  k} =
\ket{0_1, 0_2, ..., 0_{k-1}, 1_k, 0_{k+1}, ..., 0_N}$. A system
prepared in this subspace remains in it. Now the detailed dynamics
depend on the local magnetic fields or exciton energies, but if
these are independent of location $i$ then the dynamics favour
quantum state transfer processes, as already mentioned. We
adopt this limit for our work here.

\begin{figure}
\includegraphics[scale=0.32]{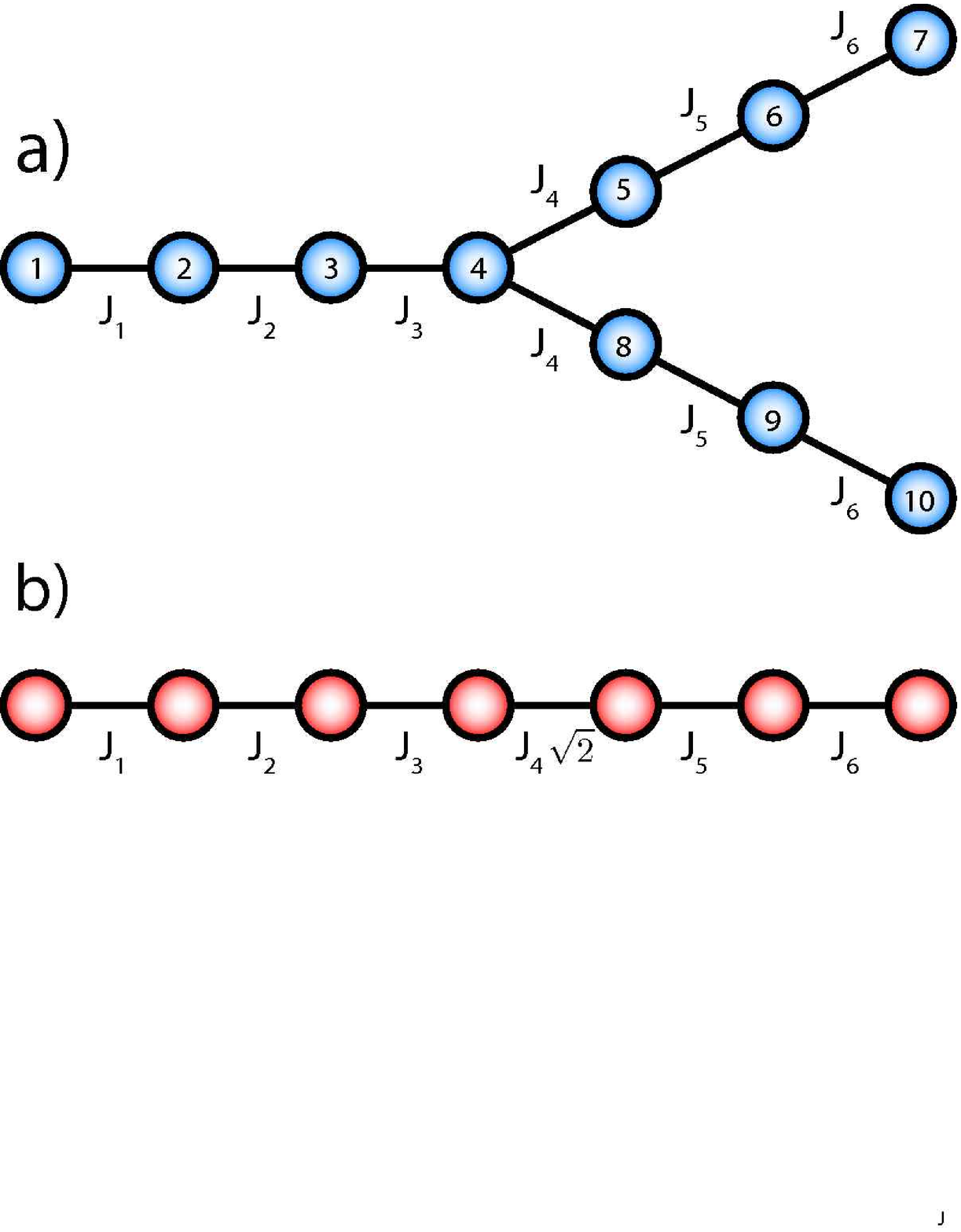}
\vspace{-3cm} \caption{(Color online) (a) Ten site Y spin chain network. Its
equivalent one-dimensional representation is shown in (b).}
\label{10site}
\end{figure}

The simplest system we consider is a Y structure---used to prepare
bi-partite entanglement from a simple initial state---labelled as 
shown in the ten site example of  Fig.~\ref{10site}.
There is a deliberate asymmetry in the coupling of the outside sites
to the hub of the Y.  We first focus on the smallest Y-example, 
which has only four sites: one `input' site, coupled to a `hub' site
with strength $J_1$, and two `output' sites, coupled to the hub
with equal strength $J_2$.
$\ket{-}\equiv2^{-\frac{1}{2}}(\ket{\mathsf
3}-\ket{\mathsf  4)}$ is an eigenstate of the system, so if we initialize
in the state $\ket{\mathsf  1}$ then $\ket{-}$
is decoupled and plays no part in the dynamics. We define the
orthogonal state $\ket{+}\equiv 2^{-\frac{1}{2}}(\ket{\mathsf
3}+\ket{\mathsf  4})$. The Hamiltonian in the $\{\ket{\mathsf 1},
\ket{\mathsf 2}, \ket{+} \}$ space is then: $H_{Y4}^\prime =
J_1\ket{\mathsf 1}\bra{\mathsf 2}+J_2\sqrt{2}\ket{\mathsf
2}\bra{\mathsf +} + H.c. .$ The system is thus equivalent to a 1D
three-site spin chain where the `output' site is the symmetric
entangled state $2^{-\frac{1}{2}}(|0\rangle_3|1\rangle_4 + |1\rangle_3|0\rangle_4)$.
This effects perfect state transfer from the input site 1 to
the entangled output so long as the couplings are equal -- i.e. if
$J_1 = J_2\sqrt{2}$.

This entanglement creation and distribution extends to spin chain
systems with longer arms. Using the method of analysis of Christandl
{\it et al.} \cite{chr05}, perfect transfer along a longer chain is
possible through the use of $N-1$ {\it unequal} couplings along an
$N$ site chain, which satisfy 
\be J_{i,i+1} = \alpha \sqrt{i(N-i)} .
\label{coupling} \ee 
where $\alpha$ is a constant that sets the size
of the interaction and $J_{i,i+1}$ is the coupling between site $i$
and site $i+1$. As detailed in Ref.~\cite{chr05}, an array of spins with equal couplings $j$ can also
be projected onto a 1D chain that has couplings $J_1, J_2\sqrt{2}$, exactly as before.
According to the general properties of a three site linear chain~\cite{chr05},
this structure allows for perfect transfer between the two extremes
so long as $J_1 = J_2\sqrt{2}$. In this case, the perfect transfer
of an excitation in node 1 is allowed across the Y-shaped structure
to the sites at the extremes 3 and 4, where it must be shared
between the two ends, giving rise to entanglement. The same
$\sqrt{2}$ hub branching factor prevents hub reflection for
propagating wave packets in divided chains of harmonic
oscillators~\cite{perales05}, as well as Gaussian wave packets of
propagating electrons or magnons~\cite{yang06}.

\begin{figure}\hspace{-5cm}
\includegraphics[scale=0.1]{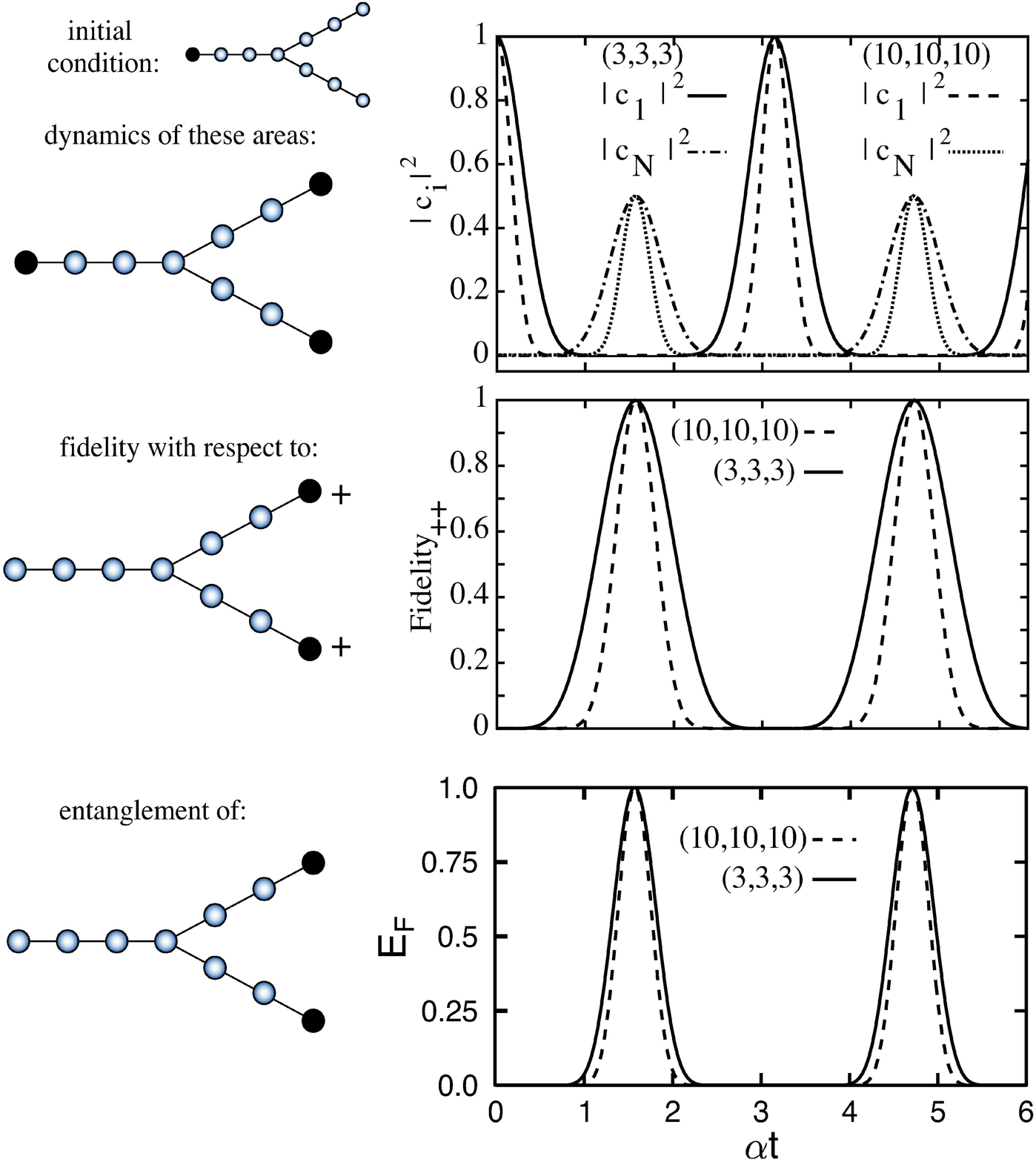}
\caption{(Color online) Results of simulations on a (3,3,3) and a (10,10,10) Y spin
chain system, with nearest neighbour couplings chosen to satisfy
 (\ref{coupling}) and the hub branching rule. Initial condition: $c_1=1$
Upper panel: $|c_1|^2$ and $|c_N|^2$ with respect to the rescaled
time $\alpha t$. Middle panel: corresponding fidelity with respect to
the output state $\ket{+}$. Lower panel: entanglement of formation of chain ends state.} \label{fig4}
\end{figure}

Extending to larger structures, we define a Y-structure by
$(l_1,l_2,l_3)$ with $l_1$ being the number of sites in the input
branch and the total site number $N=l_1+l_2+l_3+1$ . As an example,
we have performed dynamical simulations using the Hamiltonian of
(\ref{Hspin}), the condition (\ref{coupling}) and the branching rule
for the couplings at the hub spin. Fig.~\ref{10site} shows how the
hub branching rules applies to the $(3, 3, 3)$ structure, where
$J_1=J_6=\alpha\sqrt{6}, J_2 = J_5 = \alpha\sqrt{10}, J_3 =
\alpha\sqrt{12}, J_4 = \alpha\sqrt{6}$ for perfect state transfer.
Fig.~\ref{fig4} shows the result of the simulations for both the
(3, 3, 3), $N=10$, and (10, 10, 10), $N=31$, structures. The initial
condition is $c_1=1$ for the input spin (and all others zero), where
$c_i$ indicates the amplitude coefficient for the state $\ket{i}$.
The upper panel of Fig.~\ref{fig4} shows the temporal evolution of
$|c_1|^2$ and the output spin $|c_N|^2$ for the two systems. We
underline that, due to symmetry, both output spins have the same
dynamics in this case, and our results show that the excitation is
completely transferred from site $1$ to the output sites at time
$\pi/2 \alpha$ and periodically returns there at regular time
intervals of $\pi/\alpha$. With respect to the rescaled time $\alpha
t$, the peak corresponding to each revival narrows as the length
$l_k$ of the branches increases. This is important when considering
an optimal practical structure for distributing entanglement, since
a narrower peak puts greater time constraints on any entanglement
extraction protocol. The middle panel of Fig.~\ref{fig4} shows the
corresponding fidelity \cite{nie00} with respect to the state
$\ket{+}$ for the two output spins.
Note that maximally entangled states can also be produced if the
length of the output branches is different from the length of the
input branch, as long as the couplings satisfy (\ref{coupling}) and
the hub branching rule. Indeed, in the limit the length of the input
branch can be reduced to zero, so the excitation is effectively made
at the centre of an odd $N$ chain. This limit may be helpful for
some physical realizations, although having an actual input branch
may be more practical in some cases.

We can quantify the entanglement at the output using the
entanglement of formation, $E_F$, which measures the number of Bell
states required to create the state of interest. For a two qubit
state it is given by:
$E_F({\rho})=h\left(\frac{1+\sqrt{1-\tau}}{2}\right)$ where $h(x) =
-x\log_2(x) - (1 - x)\log_2(1 - x)$ is the Shannon entropy
function. $\tau$ is the ``tangle'' or ``concurrence'' squared:
$\tau={\cal
C}^{2}=\left[\max\{\lambda_1-\lambda_2-\lambda_3-\lambda_4,0\}\right]^{2}$.
The $\lambda$'s are the square roots of the eigenvalues, in
decreasing order, of the matrix ${\rho} \tilde{{\rho}} =
{\rho}\;\sigma_{y}^{A} \otimes \sigma_{y}^{B} {\rho}^{*} \sigma_y^A
\otimes \sigma_{y}^{B}$, where ${\rho}^{*}$ denotes the complex
conjugation of ${\rho}$ in the computational basis $\ket{00},
\ket{01}, \ket{10}, \ket{11}$~\cite{munro01}. $E_F$ between qubits
$n_2=7$ and $n_3=10$ of the (3, 3, 3) structure, and $n_2=21$ and $n_3=31$ of the (10, 10, 10) structure are shown in the lower panel of
Fig.~\ref{fig4}, as a function of time following initialization of
the excitation on site 1. As expected, a maximally
entangled state is obtained after a time $\pi/2\alpha$ and at
intervals of $\pi/\alpha$ thereafter.

To provide a resource, having generated and distributed entanglement
using a spin chain system, it should be isolated or frozen in some
way. Clearly, as illustrated in Fig.~\ref{fig4}, the arrival time
at the output relative to the preparation time at the input is
known. At this point, particularly if a number of such states are to
be purified before use, the dynamics have to be halted. This could
be done by extracting an entangled state (e.g. with swap operations)
from the spin chain system and transferring it to a storage system
of qubits, or to members of spatially separated quantum registers or
processors. (see e.g. \cite{spi06}).
Another possibility is to physically isolate the two nodes at the
end of the chain, after the entanglement has formed there.   If fast
local control of the couplings is possible, then the state could
indeed be frozen at the ends of the output branches. To achieve this
the couplings have to be switched, fast on a timescale set by the
chain dynamics illustrated in Fig.~\ref{fig4}, simultaneously on
both output branches.
Both of these methods present obvious difficulties, and are rather cumbersome, so we now discuss a new approach to freezing separated
entanglement, which could be very effective for some realizations of
the spin chain systems.

A very simple action that can be taken at
the end-chain entanglement arrival time is to apply a phase flip to just one
of the two output spins, transforming the state to $\ket{-}$. For
extended chains this is not an eigenstate. However, from the
symmetry of the Hamiltonian, its subsequent dynamics cannot involve the hub or
the input branch, so the subsequent `revival time' of the state
$\ket{-}$ at the output is reduced. This in itself might be useful,
but if the structure is modified slightly, coordinated action can provide {\it complete} freezing.

\begin{figure}
\includegraphics[scale=0.1]{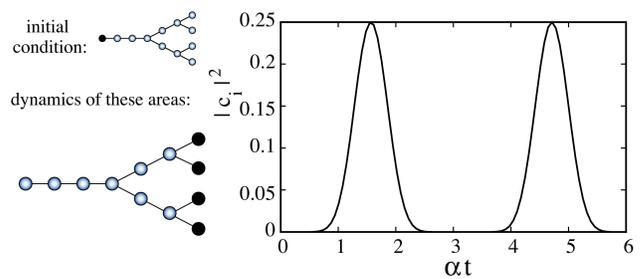}
\caption{(Color online) Dynamics of $|c_{n_i}|^2$  (for the output branch end
spins) versus the rescaled time $\alpha t$, for the bifurcation
structure shown. This follows the initialization condition $c_1 =
1$. } \label{figbifent}
\end{figure}

The required device has a Y structure, with the ultimate spin at
each output branch replaced by a bifurcation into two final spins.
These both couple to the penultimate spin with a strength reduced by
$1/\sqrt{2}$, compared to the single coupling that they replace. In
terms of the four end spins of the full device, the natural dynamics
will generate a state of the form
$\ket{\psi_s}=\frac{1}{2}(\ket{0,0,0,1}+\ket{0,0,1,0}+\ket{0,1,0,0}+\ket{1,0,0,0})$.
The structure and dynamics are shown in Fig.~\ref{figbifent}, where
we plot the branch end spin probabilities  $|c_{n_i}|^2$ as a
function of rescaled time $\alpha t$. If at a probability peak a
phase flip is applied to one spin out of each pair, a state of the
form
$\ket{\psi_a}=\frac{1}{2}(\ket{0,0,0,1}-\ket{0,0,1,0}+\ket{0,1,0,0}-\ket{1,0,0,0})$
results. This is an eigenstate of the system and the entanglement is
thus frozen. Although four spins are involved, the spatial
separation is between two pairs of spins. Each pair can
be viewed as a storage buffer for one qubit. The system contains {\it
spatially separated and stored bipartite entanglement}, which could
be released for future use by single qubit operations and/or
coupling to other systems at the branch ends.

\begin{figure}
\includegraphics[scale=0.47]{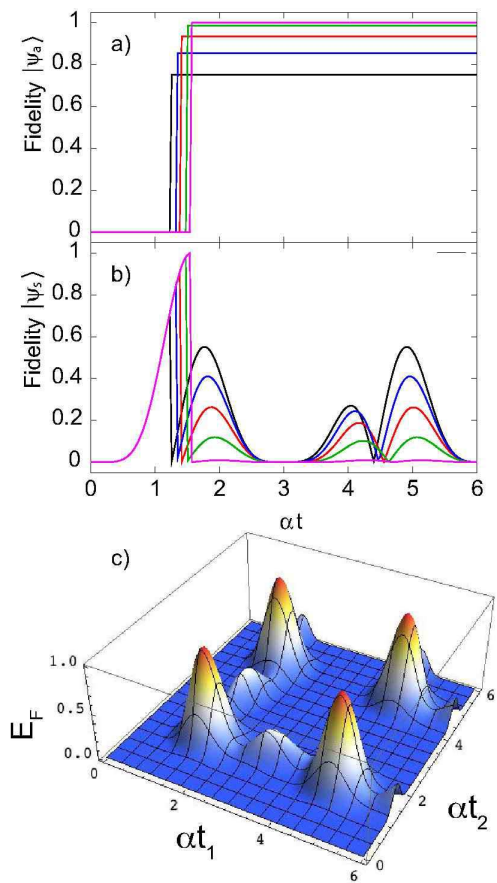}
\caption{(Color online) Behaviour of the frozen entanglement as a function of phase
flip timing, for the system of Fig. \ref{figbifent}. (a) Time
dependence of the fidelity against $\ket{\psi_a}$ for five different
flip times, assumed to be simultaneous with each other. (b) Time dependence of the fidelity
against $\ket{\psi_s}$ for the same five flip times.
 (c) Frozen bipartite entanglement of formation $E_F$ as a
function of the times $t_1$ and $t_2$ at which the phase flips are applied.} \label{figfreeze}
\end{figure}

To further quantify the entanglement freezing, we investigate the
dependence on the timing of the coordinated phase flips. The upper
panel of Fig.~\ref{figfreeze} shows the fidelity of the state as a
function of time, measured against the desired frozen state
$\ket{\psi_a}$, for five different choices of the (simultaneous) phase flip timing.
This shows that even if the timing is not perfect a high proportion of
the state is frozen. The middle panel shows the subsequent dynamics
of the remains of the state, for the same five examples of phase
flip timing, through plots of the fidelity against the un-flipped
state $\ket{\psi_s}$. These amplitudes continue to propagate
periodically through the spin chain system, with an overall
normalization set by the timing of the phase flips. The lower panel
quantifies the amount of bipartite entanglement frozen, showing the
entanglement of formation $E_F$ as a function of the times $t_1$ and $t_2$ of each pulse. The encoded basis used for these
calculations uses states of the pairs of spins at each chain end and
takes the following form: $\ket{0_L} \equiv \ket{00}$ and $\ket{1_L}
\equiv 2^{-1/2}(\ket{01}-\ket{10})$. Clearly if the phase flips are
fast compared to the system dynamics, and can be
timed sufficiently accurately and simultaneously compared to the period of these dynamics, it is
possible to freeze a high quality spatially separated Bell state,
encoded in the two pairs of spins.

To conclude, we have shown how entanglement---created  and
distributed in branching spin chain systems---can be frozen, encoded
into pairs of spins located at the ends of Y networks. Such
distributed  entanglement provides a useful resource, for example
for teleportation or distributed quantum processing.
In contrast to the use of spin chains to propagate quantum states
from one place to another with as high a fidelity as possible, we see
some advantage in building up a high fidelity entangled
resource off line. Real systems, with their inevitable
imperfections, will almost certainly degrade transmission
fidelities, even if in principle these approach unity. Certainly,
with the `off-line' resource approach, purification~\cite{pur96}
could be applied to build up a higher fidelity resource than can be
achieved by direct transmission. This can then be used to transfer
quantum states or some form of quantum communication. In effect, the
concept of a quantum repeater \cite{repeater98} could be employed in
a solid state, spin chain scenario. The key
ingredient---entanglement distribution and freezing---results from
the basic dynamics of the branched spin chain systems, simply
prepared initial states and coordinated single qubit operations.
 This opens new possibilities for quantum communication in solid state
systems, especially as fabrication or creation of solid state
systems that can operate as spin chains continues to progress.

B. W. L. acknowledges support from the Royal Society and QIPIRC (No. GR/S82176/01).

\end{document}